\documentclass[aps,prd,reprint,nofootinbib,superscriptaddress,floatfix]{revtex4-2}

\usepackage{amsmath,amssymb,bm,graphicx}
\usepackage[colorlinks=true,citecolor=blue,linkcolor=blue,urlcolor=blue]{hyperref}

\newcommand{\bx}{\boldsymbol{x}}
\newcommand{\ba}{\boldsymbol{a}}
\newcommand{\bh}{\boldsymbol{h}}
\newcommand{\bv}{\boldsymbol{v}}
\newcommand{\bQ}{\boldsymbol{Q}}
\newcommand{\bq}{\boldsymbol{q}}
\newcommand{\cL}{\mathcal{L}}
\newcommand{\dd}{\mathrm{d}}

\begin{document}

\title{Revisiting the Frolov--Zelnikov solution for Majumdar--Papapetrou black holes}

\author{Jongheon Baek}
\email{bjh6941@yonsei.ac.kr}
\affiliation{Department of Physics, Yonsei University, Seoul 03722, South Korea}

\author{Gihwan Nam}
\email{namgh@yonsei.ac.kr}
\affiliation{Department of Physics, Yonsei University, Seoul 03722, South Korea}

\date{September 19, 2026}

\begin{abstract}
We revisit Frolov and Zelnikov's electrostatic Green function for a test charge in a multi-center Majumdar--Papapetrou spacetime, correcting its horizon-charge condition and extending the result to higher dimensions.
In four dimensions, we derive the exact Dirichlet-to-Neumann map for a minimally coupled spectator Maxwell field with independently prescribed horizon potentials. The horizon flux charges obey a capacitance matrix with entries $C_{AA}=M_A+\sum_{B\ne A}M_AM_B/R_{AB}$ and $C_{AB}=-M_AM_B/R_{AB}$, where $M_A$ are center masses and $R_{AB}$ their separations. This matrix is a positive diagonal term plus a weighted graph Laplacian, yielding an exact common-mode/differential-mode decomposition of the field energy.
The grounded Dirichlet Green function partitions a point charge's flux among infinity and the horizons through harmonic measures. We show that the diagonal condition imposed by Frolov and Zelnikov [Eq.~(4.14)] preserves only the total horizon charge, allowing redistribution between individual horizons. A corrected symmetric Green function preserving each horizon charge instead involves the inverse capacitance matrix.
For two centers, the correction produces a finite, negative-semidefinite self-energy shift and an exact self-force difference, without additional ultraviolet subtraction. Finally, we extend the capacitance matrix, flux partition, and fixed-charge Green function to $D=n+3$ dimensions, where the graph weights contain the universal factor $\pi^{n/2}/\Gamma(n/2)$.
\end{abstract}

\maketitle

\section{Introduction}

The electrostatic field of a charge outside a black hole has served as a useful test of boundary conditions in curved spacetime since the calculations of Copson and Linet~\cite{Copson1928,Linet1976}.
The Schwarzschild field, its horizon polarization, and its self-force were subsequently clarified in Refs.~\cite{HanniRuffini1973,SmithWill1980,FrolovZelnikov1982}, while the Reissner--Nordstr\"om problem was solved in closed form by L\'eaut\'e and Linet~\cite{LeauteLinet1976}.
Much of this solvability can be organized through the familiar equivalence between Maxwell theory on a static geometry and electrodynamics in an effective medium~\cite{LandauLifshitz,Plebanski1960,deFelice1971}.
The same equivalence underlies optical analogues of Schwarzschild geometry~\cite{Genov2009,ChenMiaoLi2010}, although the boundary observable studied below is a strictly electrostatic one.

The one-center problem does not reveal all of the boundary data available when the event horizon is disconnected.
Majumdar--Papapetrou geometries~\cite{Majumdar1947,Papapetrou1947,HartleHawking1972} contain an arbitrary number of extremal charged black holes in static equilibrium, so that each horizon supplies an independent equipotential component.
Frolov and Zelnikov found exact scalar and electromagnetic point-source Green functions on these backgrounds, including their higher-dimensional generalization~\cite{FrolovZelnikov2012}.
Their published Maxwell kernel fixes a particular diagonal homogeneous completion through the coefficient condition in their Eq.~(4.14), which they interpret as leaving the black-hole charges unchanged.
What has remained implicit is the full boundary response: given the potentials of all horizon components, what charges flow through them, and which homogeneous completion corresponds to fixing each charge separately?
Although the membrane paradigm supplies the local conductor interpretation of a horizon~\cite{MacdonaldSuen1985,PriceThorne1986,MembraneParadigm}, we are not aware of a previous closed-form Dirichlet-to-Neumann matrix resolving the individual horizon components of a multicenter MP exterior, or of the associated finite-rank conversion between fixed-potential and fixed-component-charge Green functions.

The present work is a correction and an extension of Ref.~\cite{FrolovZelnikov2012} in this respect.
We first solve the above problem in four spacetime dimensions.
The essential reduction is short, but it produces quantities absent from the one-center analysis: an exact $N\times N$ horizon capacitance matrix, its network-energy representation, a closed flux-partition rule, and a finite-rank conversion among the relevant Green functions.
The distinction among these ensembles is physical because a vanishing sum of horizon fluxes does not imply a vanishing flux through each disconnected component.
For the standard multicenter Green function of Ref.~\cite{FrolovZelnikov2012}, a direct Gauss-flux calculation gives zero total perturbation of the horizon charge but a generically nonzero transfer between individual horizons.
So, when there is more than one center, the condition in their Eq.~(4.14) does not keep the charge of each black hole, and we give the Green function that does.
For a binary, the same finite-rank difference produces an exact shift of the renormalized electrostatic self-energy and of the force on a held charge, connecting the horizon ensemble directly to a local observable~\cite{PoissonPoundVega2011,CasalsPoissonVega2012,FrolovZelnikovAnomaly2012}.

We work in Gaussian units and set $G=c=1$.
The perturbing potential is a minimally coupled spectator Maxwell field on a fixed Majumdar--Papapetrou geometry.
If it is identified instead with a perturbation of the same Einstein--Maxwell field that supports the background, its linear stress tensor couples to metric perturbations; that coupled problem, studied in related settings such as Ref.~\cite{BiniGeralicoRuffini2007}, lies outside the test-field calculation presented here.
The membrane interpretation of horizon charge and potential provides useful intuition~\cite{MembraneParadigm}, but every response coefficient below is defined invariantly by Maxwell flux.

This paper is organized as follows.
In Secs.~\ref{sec:static} and \ref{sec:mpreduction}, we summarize the electrostatic equation on a static spacetime and its reduction on the Majumdar--Papapetrou background.
In Sec.~\ref{sec:capacitance}, we derive the horizon capacitance matrix, which is the basic quantity of this work; all the later results are obtained from it.
In Sec.~\ref{sec:grounded}, we give the grounded Green function and the flux partition of a point charge.
In Sec.~\ref{sec:ensembles}, we compare the Green function of Ref.~\cite{FrolovZelnikov2012} with the one that preserves the charge of every horizon; this section contains the correction.
Sections~\ref{sec:binary} and \ref{sec:selfenergy} show its consequences for a binary, namely the charge transfer between the horizons, the self-energy, and the self-force.
The one-center case is discussed in Sec.~\ref{sec:onecenter}, and Sec.~\ref{sec:conclusion} is devoted to the conclusion.
The extension to higher dimensions, which follows Ref.~\cite{FrolovZelnikov2012}, and the detailed comparison with their notation are given in the Appendices.

\section{Electrostatic data on a static spacetime}
\label{sec:static}

Consider a static metric with lapse $N$ and spatial metric $h_{ij}$,
\begin{equation}
 \dd s^2=-N^2\dd t^2+h_{ij}\dd x^i\dd x^j,
 \qquad A=-\Phi(\bx)\,\dd t .
 \label{eq:staticmetric}
\end{equation}
Maxwell's equation reduces on a time slice to
\begin{equation}
 -\partial_i\!\left(\frac{\sqrt{h}}{N}h^{ij}\partial_j\Phi\right)
 =4\pi\rho ,
 \label{eq:generaloperator}
\end{equation}
where $\rho$ denotes the coordinate charge density appearing after the time integration.
The lapse enters through $N^{-1}$ as the electrostatic constitutive factor on the curved spatial slice, which is the static specialization of the effective-medium correspondence~\cite{LandauLifshitz,Plebanski1960,deFelice1971}.

For a source-free region bounded by equipotential components, the Killing energy is
\begin{equation}
 E=\frac{1}{8\pi}\int\dd^3x\,
 \frac{\sqrt{h}}{N}h^{ij}\partial_i\Phi\partial_j\Phi .
 \label{eq:energygeneral}
\end{equation}
Varying this functional with fixed boundary values gives Eq.~\eqref{eq:generaloperator}.
The normal Maxwell flux through a connected inner boundary defines its charge, with the sign chosen positive when electric flux emerges from that boundary into the exterior.
The flux definition continues smoothly to a Killing horizon when it is evaluated on a family of enclosing surfaces before the horizon limit is taken.

\section{Majumdar--Papapetrou reduction}
\label{sec:mpreduction}

The four-dimensional Majumdar--Papapetrou line element is
\begin{align}
 \dd s^2&=-U^{-2}\dd t^2+U^2\dd\bx^2,
 \nonumber\\
 U(\bx)&=1+\sum_{A=1}^{N}\frac{M_A}{r_A},
 \qquad r_A=|\bx-\bx_A| .
 \label{eq:mpmetric}
\end{align}
The points $\bx_A$ are degenerate horizon ends rather than material points in the physical spatial geometry.
Their Euclidean coordinate separations
$R_{AB}=|\bx_A-\bx_B|$ are moduli of the solution and should not be confused with finite proper distances along the infinitely long extremal throats.

Equation~\eqref{eq:generaloperator} becomes
\begin{equation}
 \cL_U\Phi\equiv-\boldsymbol{\nabla}\!\cdot
 \bigl(U^2\boldsymbol{\nabla}\Phi\bigr)=4\pi\rho .
 \label{eq:mpoperator}
\end{equation}
Writing $\Psi=U\Phi$ gives the exact identity
\begin{equation}
 \cL_U\!\left(\frac{\Psi}{U}\right)
 =-U\nabla^2\Psi+\Psi\nabla^2U .
 \label{eq:doobidentity}
\end{equation}
Since $U$ is flat-harmonic away from the horizon punctures, every source-free electrostatic solution is generated locally by a flat-harmonic function $\Psi$.
The multiplicative reduction is a Doob transform~\cite{Doob1957}; it is unrelated to either the Maxwell gauge freedom or the harmonic-coordinate gauge of general relativity.

The functions
\begin{equation}
 h_A(\bx)=\frac{M_A}{r_AU(\bx)},
 \qquad h_\infty(\bx)=\frac{1}{U(\bx)}
 \label{eq:harmonicmeasures}
\end{equation}
obey the source-free equation and furnish the harmonic measures of the $N+1$ asymptotic ends.
Their boundary values and partition identity are
\begin{align}
 h_A|_{H_B}&=\delta_{AB}, & h_A|_\infty&=0,
 \nonumber\\
 h_\infty|_{H_A}&=0, & h_\infty|_\infty&=1,
 \qquad h_\infty+\sum_Ah_A=1 .
 \label{eq:hboundary}
\end{align}
Near $H_A$, the allowed flat-harmonic monopole in $\Psi$ produces a constant limiting value of $\Phi$, whereas a pole with angular momentum $\ell\geq1$ would make the Maxwell invariant singular.
Each regular connected horizon is consequently equipotential, but distinct components may carry distinct constants.
Gauge patches can be used if one also requires a regular potential one-form across several horizons; the field strength and the flux data used below are globally unambiguous.

\section{Exact horizon capacitance matrix}
\label{sec:capacitance}

Let the potential approach $V_A$ on $H_A$ and $V_\infty$ at spatial infinity.
The unique finite-energy source-free solution is
\begin{align}
 \Phi_0(\bx)&=V_\infty h_\infty(\bx)+\sum_AV_Ah_A(\bx)
 \nonumber\\
 &=\frac{V_\infty+\sum_A M_AV_A/r_A}{U(\bx)} .
\label{eq:exactpotential}
\end{align}
Uniqueness follows by applying the energy identity to the difference of two solutions with vanishing boundary data.
It is convenient to use potentials relative to infinity, $v_A=V_A-V_\infty$.

The charge of the $A$th horizon is defined by
\begin{equation}
 Q_A=-\frac{1}{4\pi}\lim_{s\to0}
 \int_{r_A=s}U^2s^2\partial_s\Phi\,\dd\Omega .
 \label{eq:horizoncharge}
\end{equation}
To evaluate it, expand the numerator and denominator of Eq.~\eqref{eq:exactpotential} near $r_A=s=0$:
\begin{align}
 U&=\frac{M_A}{s}+u_A+O(s),
 &u_A&=1+\sum_{B\ne A}\frac{M_B}{R_{AB}},
 \nonumber\\
 U\Phi_0&=\frac{M_AV_A}{s}+w_A+O(s),
 &w_A&=V_\infty+\sum_{B\ne A}\frac{M_BV_B}{R_{AB}} .
 \label{eq:nearexpansion}
\end{align}
The finite coefficient of $\partial_s\Phi_0$ then gives
\begin{equation}
 Q_A=M_Av_A+\sum_{B\ne A}\frac{M_AM_B}{R_{AB}}(v_A-v_B).
 \label{eq:chargeexplicit}
\end{equation}
Accordingly, $\bQ=C\bv$ with
\begin{align}
 C_{AA}&=M_A+\sum_{B\ne A}\frac{M_AM_B}{R_{AB}},
 \nonumber\\
 C_{AB}&=-\frac{M_AM_B}{R_{AB}}\qquad(A\ne B).
 \label{eq:capmatrix}
\end{align}
Equation~\eqref{eq:capmatrix} is the exact Dirichlet-to-Neumann map for arbitrary masses, positions, and number of horizons, with the terminology following the standard capacitance-matrix convention~\cite{SmolicKlajn2021}.
Here capacitance refers only to the response of a test Maxwell field on the fixed background, rather than a thermodynamic capacitance of the self-gravitating solution.

The structure of $C$ is particularly transparent if
$D_M=\operatorname{diag}(M_1,\ldots,M_N)$ and $L_w$ is the graph Laplacian with edge weights $w_{AB}=M_AM_B/R_{AB}$.
Then
\begin{align}
 C&=D_M+L_w,
 \nonumber\\
 \bv^{T}C\bv&=\sum_AM_Av_A^2
 +\sum_{A<B}\frac{M_AM_B}{R_{AB}}(v_A-v_B)^2 .
\label{eq:graphform}
\end{align}
The decomposition makes the matrix symmetric and strictly positive for positive $M_A$, without an appeal to a separate abstract reciprocity theorem.
The field energy reduces to the boundary expression
\begin{equation}
 E=\frac12\bv^{T}C\bv
 =\frac12\sum_AM_Av_A^2+
 \frac12\sum_{A<B}\frac{M_AM_B}{R_{AB}}(v_A-v_B)^2 .
 \label{eq:networkenergy}
\end{equation}
Every horizon is connected to infinity by a weight $M_A$, while every pair of horizons is connected by a weight $M_AM_B/R_{AB}$.
The complete-graph network is an exact representation of the curved-spacetime electrostatic energy, not a large-separation circuit approximation.

The row sums satisfy
\begin{equation}
 \sum_BC_{AB}=M_A,
 \qquad
 \sum_AQ_A=\sum_AM_Av_A .
 \label{eq:rowsums}
\end{equation}
An extended matrix including infinity has entries
$C_{A\infty}=C_{\infty A}=-M_A$ and
$C_{\infty\infty}=\sum_AM_A$; every row of that matrix sums to zero, as required by invariance under a common potential shift.
The off-diagonal entries also have an immediate induction meaning: when only horizon $A$ is raised above infinity, every other horizon acquires a negative perturbation charge proportional to $M_AM_B/R_{AB}$.

\begin{figure*}[t]
 \includegraphics[width=\textwidth]{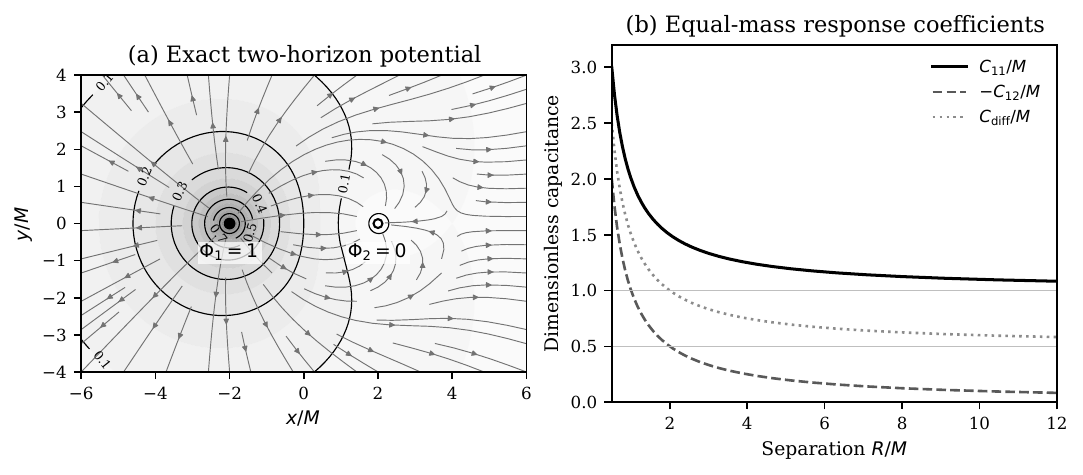}
 \caption{Exact fixed-potential response of an equal-mass binary with $R=4M$.
 (a) Equipotential contours and electric-flux streamlines for $(v_1,v_2)=(1,0)$; the punctures represent the two extremal horizon ends.
 (b) The diagonal element, mutual coefficient, and differential capacitance as functions of coordinate separation.
 All curves follow directly from Eqs.~\eqref{eq:exactpotential} and \eqref{eq:capmatrix}.}
 \label{fig:capacitance}
\end{figure*}

\section{Grounded Green function and flux partition}
\label{sec:grounded}

Place a point charge at $\ba$, away from the horizons, and write
$U_a=U(\ba)$ and $a_A=|\ba-\bx_A|$.
The Green function that vanishes on every horizon and at infinity is
\begin{equation}
 G_D(\bx,\ba)=
 \frac{1}{U(\bx)U_a|\bx-\ba|},
 \qquad
 \cL_U G_D=4\pi\delta^3(\bx-\ba).
 \label{eq:dirichletgreen}
\end{equation}
The kernel is symmetric, has the required local singularity
$[U_a^2|\bx-\ba|]^{-1}$, and its boundary values follow directly from the divergent behavior of $U$ at each puncture.
The transformation in Eq.~\eqref{eq:doobidentity} supplies a direct proof: $U(\bx)G_D=1/[U_a|\bx-\ba|]$ is the flat Coulomb kernel with the normalization needed at its source.

For a charge $q$, the horizon flux obtained from Eq.~\eqref{eq:horizoncharge} is
\begin{equation}
 Q_A^{D}=-q h_A(\ba)
 =-q\frac{M_A}{U_a a_A},
 \qquad
 q_\infty=\frac{q}{U_a}.
 \label{eq:inducedcharges}
\end{equation}
Hence the nonnegative fractions of source flux absorbed by each grounded horizon and transmitted to infinity are
\begin{equation}
 p_A=\frac{M_A}{U_a a_A},
 \qquad p_\infty=\frac{1}{U_a},
 \qquad p_\infty+\sum_Ap_A=1 .
 \label{eq:fluxpartition}
\end{equation}
The partition of unity in Eq.~\eqref{eq:hboundary} becomes an exact flux-splitting rule for the point charge.
The result remains elementary even when the horizon geometry and the field lines are strongly distorted by neighboring centers.

With prescribed horizon potentials and several external charges $q_s$ at $\ba_s$, linearity gives the complete response
\begin{align}
 \Phi(\bx)&=V_\infty+\sum_Av_Ah_A(\bx)
 +\sum_s q_sG_D(\bx,\ba_s),
 \nonumber\\
 Q_A&=\sum_BC_{AB}v_B-\sum_sq_sh_A(\ba_s).
 \label{eq:masterresponse}
\end{align}
Equation~\eqref{eq:masterresponse} separates the voltage-controlled horizon response from the induced charge of the external sources.

\section{Fixed-potential and fixed-charge ensembles}
\label{sec:ensembles}

The local source singularity does not determine a static Green function until its horizon data are specified.
The grounded kernel $G_D$ describes the fixed-potential ensemble, whereas adding a homogeneous combination $\sum_A\alpha_Ah_A(\bx)$ changes the vector of horizon potentials by $\boldsymbol{\alpha}$ and the horizon charges by $C\boldsymbol{\alpha}$.
The resulting conversion between boundary ensembles is finite dimensional.

In the present four-dimensional normalization, the electromagnetic kernel of Frolov and Zelnikov~\cite{FrolovZelnikov2012} can be written as
\begin{equation}
 G_{\rm FZ}(\bx,\ba)=G_D(\bx,\ba)
 +\sum_A\frac{h_A(\bx)h_A(\ba)}{M_A} .
 \label{eq:fzgreen}
\end{equation}
This is precisely their published Eq.~(4.17) after converting the sign of the electrostatic potential and the normalization of the point source; Appendix~\ref{app:fzmapping} gives the conversion from their $D$-dimensional formula.
Frolov and Zelnikov obtained this expression from the diagonal coefficient condition in their Eq.~(4.14), which they interpreted as removing the additional center poles while leaving the black-hole charges unchanged.
For one center this interpretation is correct, but for several centers finite cross terms survive in the flux through each puncture.
A later biconformal treatment likewise fixes a homogeneous zero mode through an aggregate flux condition~\cite{FrolovZelnikov2015}.

The grounded part contributes $-h_A(\ba)$ to the $A$th horizon charge per unit source charge.
Applying the full matrix~\eqref{eq:capmatrix} to the horizon values $h_B(\ba)/M_B$ of the added term gives
\begin{align}
 \delta Q_A^{\rm FZ}
 &= -h_A(\ba)+\sum_BC_{AB}\frac{h_B(\ba)}{M_B}
 \nonumber\\
 &=\frac{M_A}{U_a}\sum_{B\ne A}\frac{M_B}{R_{AB}}
 \left(\frac{1}{a_A}-\frac{1}{a_B}\right).
 \label{eq:fzflux}
\end{align}
The pairwise antisymmetry of the second line implies
\begin{equation}
 \sum_A\delta Q_A^{\rm FZ}=0,
 \label{eq:fztotal}
\end{equation}
but the individual terms are generally nonzero.
Thus Eq.~\eqref{eq:fzgreen} is a valid homogeneous completion of the exterior Green function, but its component-resolved boundary interpretation differs from the one-center case: it preserves the total horizon charge while allowing charge to move between disconnected horizons.
Componentwise preservation occurs only for one center, for a source equidistant from all interacting centers, or in other configurations where Eq.~\eqref{eq:fzflux} happens to vanish.
Therefore, for more than one center, the interpretation of the condition~(4.14) in Ref.~\cite{FrolovZelnikov2012} should be corrected: this condition fixes the sum of the horizon charges, not the charge of each black hole.

To hold every horizon charge fixed, the required change of horizon potential is obtained from the full inverse response matrix.
The corresponding symmetric Green function is
\begin{equation}
 G_Q(\bx,\ba)=G_D(\bx,\ba)
 +\bh(\bx)^TC^{-1}\bh(\ba)
 \label{eq:fixedchargegreen}
\end{equation}
with $\bh=(h_1,\ldots,h_N)^T$.
Indeed, the homogeneous term changes the charge vector by $C C^{-1}\bh(\ba)=\bh(\ba)$ and exactly cancels the induced vector $-\bh(\ba)$ of the grounded solution.
Moreover, $C\boldsymbol{1}=(M_1,\ldots,M_N)^T$ and Eq.~\eqref{eq:hboundary} show that the asymptotic $1/r$ coefficient of $G_Q$ is unity, so the complete unit source flux reaches infinity when every horizon flux is held fixed.

Equations~\eqref{eq:dirichletgreen}, \eqref{eq:fzgreen}, and \eqref{eq:fixedchargegreen} encode three distinct pieces of boundary information.
They prescribe, respectively, all horizon potentials, only the aggregate horizon charge, and every individual horizon charge.
Their differences are finite-rank kernels built solely from the horizon harmonic measures.

\section{Equal-mass binary}
\label{sec:binary}

For two holes of equal mass $M$ and coordinate separation $R$, the response matrix is
\begin{equation}
 C=M
 \begin{pmatrix}
 1+M/R & -M/R\\
 -M/R & 1+M/R
 \end{pmatrix} .
 \label{eq:binarymatrix}
\end{equation}
The common and differential eigenvectors $(1,1)$ and $(1,-1)$ have eigenvalues
\begin{equation}
 \lambda_{\rm com}=M,
 \qquad
 \lambda_{\rm dif}=M\left(1+\frac{2M}{R}\right).
 \label{eq:binaryeigenvalues}
\end{equation}
The common mode is insensitive to the separation because the pair term in Eq.~\eqref{eq:networkenergy} depends only on potential differences.
If the binary is driven antisymmetrically, $v_1=\Delta V/2$ and $v_2=-\Delta V/2$, the charge on the first horizon is $Q_1=C_{\rm diff}\Delta V$ with
\begin{equation}
 C_{\rm diff}=\frac{M}{2}+\frac{M^2}{R} .
 \label{eq:differentialcapacitance}
\end{equation}
Figure~\ref{fig:capacitance} displays both the exact field and these response coefficients.
The formal coalescence limit needs some care.
As $R\to0$, $C_{11}$ and the differential eigenvalue diverge, whereas the common-mode response remains $(Q_1+Q_2)/v=2M$, equal to the capacitance of one extremal hole of mass $2M$.
For a source kept at a fixed exterior position, the differential harmonic measure is $O(R)$, so $G_Q-G_{\rm FZ}=O(R^2)$.
The exterior common sector therefore has a smooth one-center limit even though the topology-changing differential sector does not.

For unequal masses, the inverse matrix is still elementary.
Writing $w=M_1M_2/R$, one finds
\begin{equation}
 C^{-1}=\frac{1}{M_1M_2+w(M_1+M_2)}
 \begin{pmatrix}M_2+w&w\\w&M_1+w\end{pmatrix}.
 \label{eq:binaryinverse}
\end{equation}
Comparison with the diagonal coefficients in Eq.~\eqref{eq:fzgreen} gives the rank-one difference
\begin{align}
 G_Q-G_{\rm FZ}
 ={}&-\frac{M_1M_2}{R+M_1+M_2}
 \nonumber\\
 &\times\left(\frac{h_1}{M_1}-\frac{h_2}{M_2}\right)_{\!\bx}
 \left(\frac{h_1}{M_1}-\frac{h_2}{M_2}\right)_{\!\ba} .
 \label{eq:rankonedifference}
\end{align}
Only the differential horizon mode is altered, consistently with the fact that both kernels preserve the same total charge.

\begin{figure*}[t]
 \includegraphics[width=\textwidth]{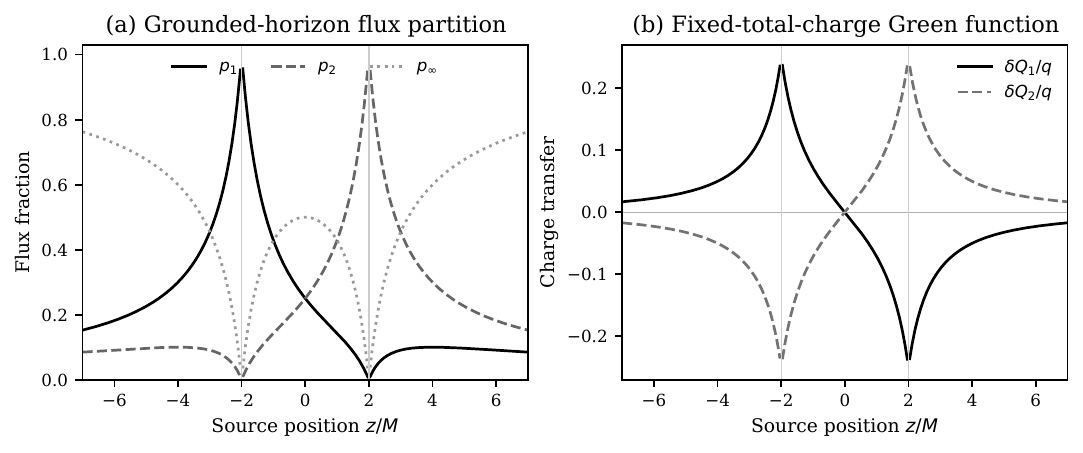}
 \caption{Point-source response for two equal masses at $z=\pm2M$.
 (a) Exact fractions of flux absorbed by either grounded horizon or transmitted to infinity as the source moves along the symmetry axis.
 (b) Individual horizon-charge shifts generated by the fixed-total-charge completion~\eqref{eq:fzgreen}; the two curves cancel at every source position, although neither is generally zero.}
 \label{fig:partition}
\end{figure*}

As a concrete check, take $M_1=M_2=M$, $R=4M$, and place a unit source where $(a_1,a_2)=(M,3M)$.
Then $U_a=7/3$, and the grounded flux fractions are
\begin{equation}
 (p_1,p_2,p_\infty)=\left(\frac{3}{7},\frac{1}{7},\frac{3}{7}\right).
 \label{eq:numericalpartition}
\end{equation}
The completion~\eqref{eq:fzgreen} produces instead the individual horizon shifts
\begin{equation}
 (\delta Q_1^{\rm FZ},\delta Q_2^{\rm FZ})
 =\left(\frac{q}{14},-\frac{q}{14}\right),
 \label{eq:numericaltransfer}
\end{equation}
which leave the total horizon charge unchanged but provide a direct counterexample to componentwise conservation.
The position dependence of both effects is shown in Fig.~\ref{fig:partition}.

\section{Self-energy and the physical ensemble}
\label{sec:selfenergy}

The finite-rank difference in Eq.~\eqref{eq:rankonedifference} changes a local observable.
For a reciprocal static Green function, the renormalized Killing self-energy of a point charge is $E_{\rm self}=q^2G_{\rm reg}(\ba,\ba)/2$~\cite{FrolovZelnikov1982,PoissonPoundVega2011,CasalsPoissonVega2012,FrolovZelnikovAnomaly2012}.
The local singular fields of $G_Q$ and $G_{\rm FZ}$ coincide, so their difference may be evaluated without any further ultraviolet subtraction.
Defining
\begin{equation}
 K=\frac{M_1M_2}{R+M_1+M_2},\qquad
 \chi(\bx)=\frac{h_1(\bx)}{M_1}-\frac{h_2(\bx)}{M_2},
 \label{eq:kchi}
\end{equation}
one obtains
\begin{equation}
 \Delta E_{\rm self}\equiv E_{\rm self}^{Q}-E_{\rm self}^{\rm FZ}
 =-\frac{q^2K}{2}\chi(\ba)^2 .
 \label{eq:selfenergyshift}
\end{equation}
The shift is negative semidefinite, vanishes when the source couples equally to both horizons, and depends only on the differential mode.

Let $u^\mu$ be the four-velocity of the static observers and $P_\mu{}^\nu=\delta_\mu{}^\nu+u_\mu u^\nu$ their spatial projector.
The corresponding difference of local self-force covectors is
\begin{equation}
 \Delta f_\mu=-\frac{1}{N}P_\mu{}^\nu\nabla_\nu\Delta E_{\rm self}
 =\frac{q^2K}{N}\chi P_\mu{}^\nu\nabla_\nu\chi .
 \label{eq:selfforceshift}
\end{equation}
In the orthonormal frame associated with the MP Cartesian coordinates, this becomes $\Delta f_{\hat i}=q^2K\chi\,\partial_i\chi$.
The factor $1/2$ in Eq.~\eqref{eq:selfenergyshift} does not halve the force, because reciprocity supplies equal derivatives from the two arguments of the coincident Green function.
The external force required to hold the particle has the opposite sign.

\begin{figure*}[t]
 \includegraphics[width=\textwidth]{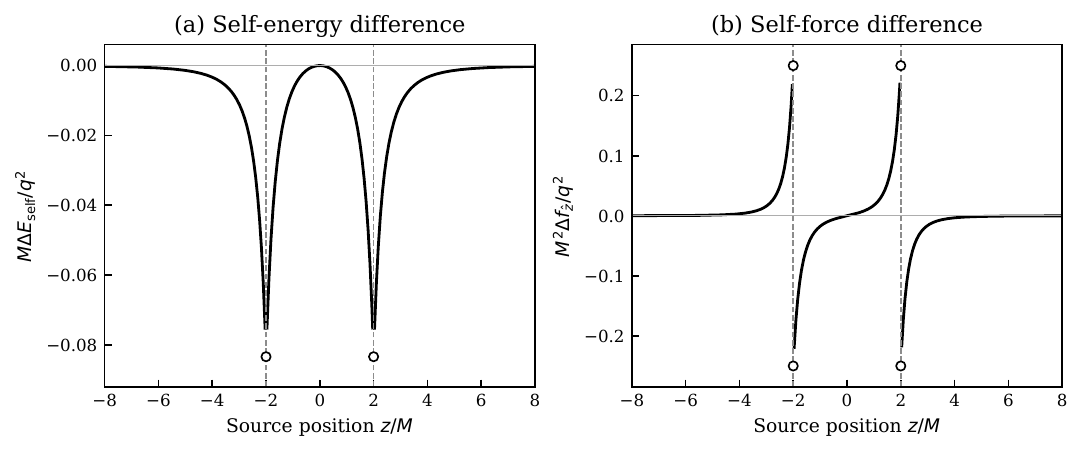}
 \caption{Ensemble-dependent self-energy and self-force for an equal-mass binary with $R=4M$.
 The curves show the component-fixed result minus Eq.~\eqref{eq:fzgreen} as the source moves on the symmetry axis.
 (a) Self-energy difference, which is finite without an additional subtraction, vanishes on the symmetry plane, and tends to $-q^2/(12M)$ at either horizon.
 (b) Orthonormal force component; the reversal at each puncture reflects the two exterior directions from which the throat is approached.}
 \label{fig:selfenergy}
\end{figure*}

For the equal-mass binary, put $\zeta=z/M$ and ${\cal R}=R/M$ along the line of centers, and define $\widehat\chi=M\chi$.
The result needed to plot both observables is
\begin{align}
 \widehat\chi(\zeta)&=
 \begin{cases}
 -\dfrac{2\zeta}{{\cal R}^2/4+{\cal R}-\zeta^2},&|\zeta|<{\cal R}/2,\\[2mm]
 -\dfrac{\operatorname{sgn}(\zeta){\cal R}}
 {\zeta^2+2|\zeta|-{\cal R}^2/4},&|\zeta|>{\cal R}/2,
 \end{cases}
 \nonumber\\
 \frac{M\Delta E_{\rm self}}{q^2}&=-\frac{\widehat\chi^2}{2({\cal R}+2)},
 \nonumber\\
 \frac{M^2\Delta f_{\hat z}}{q^2}&=
 \frac{\widehat\chi\widehat\chi'}{{\cal R}+2}.
 \label{eq:axisobservables}
\end{align}
For $R=4M$, the energy shift approaches $-q^2/(12M)$ at either horizon and decays as $|z|^{-4}$ at infinity; the force difference decays as $|z|^{-5}$.
Figure~\ref{fig:selfenergy}(b) also shows that the force branch changes direction when the same throat is approached from its two exterior sides.

Which boundary condition is physical depends on the Maxwell sector.
For an independent spectator $U(1)$ field, its horizon fluxes are not the charges supporting the MP geometry, and extremality does not select an ensemble.
If the perturbation is instead identified with the supporting Einstein--Maxwell field while the masses and geometry are held fixed, $\delta Q_A=0$ is the natural electromagnetic boundary condition, and $G_Q$ supplies this sector.
This statement does not mean that every nonzero $\delta Q_A$ by itself makes a horizon superextremal.
The supporting field changes the stress tensor at linear order and generally requires a metric perturbation, while variations tangent to the MP family obey $\delta M_A=\operatorname{sgn}(Q_A)\delta Q_A$.
A complete self-force calculation in that sector must therefore solve the coupled linearized Einstein--Maxwell problem; the result above isolates the boundary-data contribution that such a calculation must use.

\section{Relation to one-center electrostatics}
\label{sec:onecenter}

For $N=1$, the graph Laplacian is absent and the capacitance reduces to $C=M$.
The two charge-controlled prescriptions then coincide because $C^{-1}=M^{-1}$, while the grounded kernel retains the induced charge $-qM/(U_aa)$.
For one center, this collapse hides the distinction between total and componentwise horizon charge in the usual extremal Reissner--Nordstr\"om problem.

The classic Copson--Linet and L\'eaut\'e--Linet solutions concern nonextremal Schwarzschild and Reissner--Nordstr\"om geometries~\cite{Copson1928,Linet1976,LeauteLinet1976}.
There the radial equation may be reduced to Legendre form, and the additional monopole is fixed by a global horizon charge condition.
With a cosmological constant the separated equation is of Heun type, but a static point-charge mode sum in Schwarzschild--de Sitter spacetime is already known~\cite{KucharPoissonVega2013}.
Those reductions are not presented here as new results.
Their role here is conceptual: a homogeneous zero mode encodes an ensemble choice, whereas the multicenter geometry promotes that one coefficient to an $N$-component response problem.

\section{Discussion and conclusion}
\label{sec:conclusion}

Disconnected extremal horizons possess a finite-dimensional electrostatic sector that is invisible in a one-center calculation.
The exact response matrix~\eqref{eq:capmatrix} shows that this sector is governed by a complete weighted graph whose vertices are the horizons and infinity.
Its diagonal part records the coupling of each throat to infinity, while its graph-Laplacian part measures differential potentials between horizons.
Because the result holds at arbitrary coordinate separation, it gives a nonperturbative boundary observable of the full Majumdar--Papapetrou family rather than a far-zone approximation.

The same matrix resolves an ambiguity in point-source Green functions.
Fixing the horizon potentials, fixing the sum of their charges, and fixing every charge are inequivalent boundary problems once the horizon has more than one component.
The direct conversion of the Frolov--Zelnikov kernel shows that its diagonal completion preserves only the aggregate horizon charge, while the inverse response matrix gives the componentwise fixed-charge kernel.
For a binary, the two choices lead to a finite negative-semidefinite self-energy difference and a position-dependent force difference, so the boundary ensemble is visible in a local observable rather than only in bookkeeping at the horizons.
The same construction survives in arbitrary dimension with a universal sphere-area normalization, although generic higher-dimensional multicenter horizons have limited differentiability.

The spectator calculation by itself does not determine the coupled gravitational self-force of a charge in the supporting gauge sector.
It does, however, identify the fixed-component-charge Maxwell boundary data required by that problem and separates them from the alternative total-charge completion.
Off-axis motion and the coupled Einstein--Maxwell perturbation are natural places where the differential horizon mode may enter a directly dynamical observable.

\section*{Data Availability}
This is a purely analytical study, and no data were created or analyzed.
All figures can be reproduced directly from the equations presented in the article and the plotting script supplied with the source.

\appendix

\section{Horizon regularity and the flux limit}
\label{app:regularity}

Near the $A$th puncture, write $s=r_A$ and expand a flat-harmonic numerator in spherical harmonics.
The two local radial branches are $s^\ell Y_{\ell m}$ and $s^{-\ell-1}Y_{\ell m}$.
Since $U\sim M_A/s$, the corresponding electrostatic potential $\Phi=\Psi/U$ behaves as $s^{\ell+1}Y_{\ell m}$ or $s^{-\ell}Y_{\ell m}$.
The singular branch is excluded for $\ell\geq1$ because $F_{\mu\nu}F^{\mu\nu}$ diverges, while the $\ell=0$ pole gives a finite constant on the horizon.
Regularity leaves the local form
\begin{equation}
 \Phi=V_A+\kappa_A s+O(s^2)
 \label{eq:localregular}
\end{equation}
up to angular terms that vanish faster than $s$.
Both the Maxwell invariant and the Killing energy density are integrable for this expansion.

For the source-free solution~\eqref{eq:exactpotential}, division of the two series in Eq.~\eqref{eq:nearexpansion} yields
\begin{equation}
 \kappa_A=\frac{M_Aw_A-M_AV_Au_A}{M_A^2} .
 \label{eq:kappaa}
\end{equation}
Because $U^2s^2\to M_A^2$, insertion into Eq.~\eqref{eq:horizoncharge} gives
$Q_A=M_AV_Au_A-M_Aw_A$, which is precisely Eq.~\eqref{eq:chargeexplicit} after subtracting $V_\infty$.

\section{Energy identity and ensemble conversion}
\label{app:energy}

Let $\Phi$ solve the source-free equation and approach $v_A$ on the horizons and zero at infinity.
Integration by parts over the exterior region, followed by the horizon limit, gives
\begin{equation}
 \frac{1}{4\pi}\int U^2|\boldsymbol{\nabla}\Phi|^2\dd^3x
 =\sum_Av_AQ_A .
 \label{eq:greenidentity}
\end{equation}
Equation~\eqref{eq:networkenergy} follows after inserting $\bQ=C\bv$ and the factor $1/2$ appropriate to electrostatic energy.
The same identity proves uniqueness when every boundary value vanishes.

For a unit point source, let $\bq_D=-\bh(\ba)$ be the induced horizon-charge vector of $G_D$.
Adding $\bh(\bx)^T\boldsymbol{\alpha}$ changes that vector to
\begin{equation}
 \bq=\bq_D+C\boldsymbol{\alpha}.
 \label{eq:ensemblelinear}
\end{equation}
Setting $\bq=0$ in Eq.~\eqref{eq:ensemblelinear} gives
$\boldsymbol{\alpha}=C^{-1}\bh(\ba)$, which proves Eq.~\eqref{eq:fixedchargegreen} without any assumption about symmetry between the centers.
For the coefficients $\alpha_A=h_A(\ba)/M_A$ used in Eq.~\eqref{eq:fzgreen}, Eq.~\eqref{eq:ensemblelinear} instead reduces to Eq.~\eqref{eq:fzflux}; summing over $A$ removes every pair term and proves Eq.~\eqref{eq:fztotal}.

\section{Higher-dimensional response}
\label{app:dimensions}

The following extension is an exterior-domain statement.
The higher-dimensional MP geometry may be written as
\begin{equation}
 \dd s^2=-U^{-2}\dd t^2+U^{2/n}\dd\bx^2,
 \qquad
 U=1+\sum_A\frac{M_A}{r_A^n}.
 \label{eq:dmetric}
\end{equation}
Here $\bx$ has $n+2$ components, $M_A$ is the pole strength in $U$, and its conversion to the ADM mass depends on the gravitational normalization.
With
\begin{align}
 \Omega_{n+1}&=\frac{2\pi^{(n+2)/2}}{\Gamma[(n+2)/2]},
 \nonumber\\
 \alpha_n&=\frac{n\Omega_{n+1}}{4\pi}
 =\frac{\pi^{n/2}}{\Gamma(n/2)},
 &\beta_n&=\alpha_n^{-1},
 \label{eq:alphabeta}
\end{align}
the flat fundamental solution obeys
$-\nabla^2[\beta_n/|\bx-\ba|^n]=4\pi\delta^{n+2}(\bx-\ba)$.
The metric factors satisfy
\begin{equation}
 \frac{\sqrt h}{N}h^{ab}=U^2\delta^{ab},
 \label{eq:dreduction}
\end{equation}
so the Maxwell operator and the Doob identity retain exactly the four-dimensional form.

The harmonic measures are $h_A=M_A/(r_A^nU)$ and $h_\infty=U^{-1}$.
Repeating the puncture-flux calculation gives
\begin{align}
 C_{AA}^{(D)}&=\alpha_n\left(M_A+
 \sum_{B\ne A}\frac{M_AM_B}{R_{AB}^n}\right),
 \nonumber\\
 C_{AB}^{(D)}&=-\alpha_n\frac{M_AM_B}{R_{AB}^n}
 \qquad(A\ne B),
 \label{eq:dcapacitance}
\end{align}
and the source-free field energy is
\begin{equation}
 E=\frac{\alpha_n}{2}\left[
 \sum_AM_Av_A^2+
 \sum_{A<B}\frac{M_AM_B}{R_{AB}^n}(v_A-v_B)^2\right].
 \label{eq:denergy}
\end{equation}
Thus positivity, the graph-Laplacian structure, and the separation into common and differential modes do not rely on $D=4$.

The grounded and componentwise fixed-charge kernels are
\begin{align}
 G_D^{(D)}(\bx,\ba)&=
 \frac{\beta_n}{U(\bx)U_a|\bx-\ba|^n},
 \nonumber\\
 G_Q^{(D)}(\bx,\ba)&=G_D^{(D)}(\bx,\ba)
 +\bh(\bx)^T[C^{(D)}]^{-1}\bh(\ba).
 \label{eq:dgreenfunctions}
\end{align}
The grounded fluxes remain $-q h_A(\ba)$ at the individual horizons and $q h_\infty(\ba)$ at infinity.
Generic multicenter MP horizons cease to be smooth in $D\geq5$~\cite{Welch1995,CandlishReall2007}; hence the puncture fluxes in Eqs.~\eqref{eq:dcapacitance}--\eqref{eq:dgreenfunctions}, although well defined from the exterior equation, should not be read without further analysis as data on a smooth horizon extension.

\section{Frolov--Zelnikov normalization and component flux}
\label{app:fzmapping}

We now translate the notation of Ref.~\cite{FrolovZelnikov2012} without inferring the boundary condition from the one-center case.
Frolov and Zelnikov set $A_0=-U^{-1}\psi$ and normalize an elementary source by $e'=1/(4\pi)$, for which their transformed equation is
\begin{equation}
 U\nabla^2\psi-\psi\nabla^2U
 =-\delta^{n+2}(\bx-\ba).
 \label{eq:fztransformed}
\end{equation}
Their $D=n+3$ ansatz, Eqs.~(4.9) and (4.14), is
\begin{align}
 \psi&=\frac{1}{n\Omega_{n+1}U_a}
 \left[\frac{1}{|\bx-\ba|^n}+b+
 \sum_A\frac{\gamma_A}{r_A^n}\right],
 \nonumber\\
 \gamma_A^{\rm FZ}&=M_A(a_A^{-n}+b).
 \label{eq:fzansatz}
\end{align}
Their asymptotic gauge choice sets $b=0$, and their published Eq.~(4.17) is
\begin{equation}
 {\cal G}_{00}^{\rm FZ}=-\frac{1}{n\Omega_{n+1}U(\bx)U_a}
 \left[\frac{1}{|\bx-\ba|^n}
 +\sum_A\frac{M_A}{r_A^na_A^n}\right].
 \label{eq:fzpublished}
\end{equation}
They use $A_0=4\pi q{\cal G}_{00}$, whereas our convention is $A_0=-\Phi=-qG$.
It follows that
\begin{equation}
 G_{\rm FZ}^{(D)}=G_D^{(D)}+
 \beta_n\sum_A\frac{h_A(\bx)h_A(\ba)}{M_A},
 \label{eq:dfzgreen}
\end{equation}
which reduces exactly to Eq.~\eqref{eq:fzgreen} when $n=1$.

The distinction between the diagonal condition in Eq.~\eqref{eq:fzansatz} and a componentwise flux condition can be seen before taking any distributional product.
Near the $A$th puncture, write
\begin{align}
 U&=\frac{M_A}{s^n}+u_A+O(s),
 &u_A&=1+\sum_{B\ne A}\frac{M_B}{R_{AB}^n},
 \nonumber\\
 S&=\frac{\gamma_A}{s^n}+s_A+O(s),
 &s_A&=a_A^{-n}+b+
 \sum_{B\ne A}\frac{\gamma_B}{R_{AB}^n},
 \label{eq:fzpunctureexpansion}
\end{align}
where $S$ denotes the square bracket in Eq.~\eqref{eq:fzansatz}.
The quotient has the local expansion
\begin{equation}
 \frac{S}{U}=\frac{\gamma_A}{M_A}
 +\frac{s^n}{M_A^2}(M_As_A-\gamma_Au_A)+O(s^{n+1}).
 \label{eq:fzquotient}
\end{equation}
The coefficient of $s^n$ fixes the component flux, so its vanishing requires
\begin{equation}
 \gamma_Au_A-M_A\sum_{B\ne A}
 \frac{\gamma_B}{R_{AB}^n}=M_A(a_A^{-n}+b).
 \label{eq:fzcoupledcondition}
\end{equation}
For $x_A=\gamma_A/M_A$ and $b=0$, this is
\begin{equation}
 x_A+\sum_{B\ne A}\frac{M_B}{R_{AB}^n}(x_A-x_B)=a_A^{-n}.
 \label{eq:fzmatrixcondition}
\end{equation}
Using Eq.~\eqref{eq:dcapacitance}, the coupled system is precisely the condition that produces $[C^{(D)}]^{-1}$ in Eq.~\eqref{eq:dgreenfunctions}.
By contrast, inserting the diagonal coefficients $\gamma_A^{\rm FZ}=M_Aa_A^{-n}$ into the same puncture flux gives
\begin{equation}
 \delta Q_A^{\rm FZ}=
 \frac{qM_A}{U_a}\sum_{B\ne A}\frac{M_B}{R_{AB}^n}
 (a_A^{-n}-a_B^{-n}),
 \label{eq:dfzflux}
\end{equation}
whose sum vanishes pairwise although its components generally do not.
The self-energy discussion of Ref.~\cite{FrolovZelnikovAnomaly2012} states after its Eq.~(2.9) that the external source has zero flux through every horizon and later imports the MP kernel in its Eqs.~(5.8)--(5.10).
Equation~\eqref{eq:dfzflux} shows that this statement holds for the sum of the connected components, but not generally for each one.
The calculation leaves the Frolov--Zelnikov exterior solution unchanged and resolves the discrepancy by replacing the diagonal prescription with Eq.~\eqref{eq:fzmatrixcondition} when componentwise Gauss constraints are required.

\end{document}